\documentclass[superscriptaddress,nofootinbib,amsmath,longbibliography,amssymb,11pt]{revtex4}
\usepackage{amsmath,amsthm}
\usepackage{graphicx}
\usepackage[colorlinks=true, linkcolor=blue, citecolor=magenta, hyperfootnotes=true]{hyperref}
\usepackage{url}

\newcommand{\nc}{\newcommand}
\nc{\ba}{\begin{eqnarray}}
\nc{\ea}{\end{eqnarray}}

\newcommand{\calT}{{\cal{T}}}
\nc{\bfx}{{\bf{x}}}
\nc{\bfk}{{\bf{k}}}
\nc{\mo}{{\theta}}
\nc{\moh}{{ \theta'}}
\nc{\mt}{{ \mu}}
\nc{\ms}{{\mu_D}}
\nc{\pval}{{p^\text{val}}}

\nc{\mhn}{\color{blue}{\bf MHN: }}
\nc{\mth}{\theta_{\text{th}}}

\begin{document}

\title{ Bayesian questions with  frequentist answers}

\author{Alan H. Guth}
\email{guth@ctp.mit.edu}
\affiliation{Department of Physics, Laboratory for Nuclear Science, and Center for Theoretical Physics, Massachusetts Institute of Technology, Cambridge, MA 02139}

\author{Mohammad Hossein Namjoo}
\email{mh.namjoo@ipm.ir}
\affiliation{School of Astronomy, Institute for Research in Fundamental Sciences (IPM), Tehran, Iran, P.O. Box 19395-5531}

\date{\today}

\begin{abstract}
The two statistical methods, namely the frequentist and the Bayesian methods, are both commonly used for probabilistic inference in many scientific situations. However, it is not straightforward to interpret the result of one approach in terms of the concepts of the other. In this paper we explore the possibility of finding a Bayesian significance for the frequentist's main object of interest, the $p$-value, which is the probability assigned to the proposition --- which we call the {\it extremity proposition} --- that a measurement will result in a value that is at least as extreme as the value that was actually obtained.  To make contact with the frequentist language, the Bayesian can choose to update probabilities based on the {\it extremity proposition}, which is weaker than the standard Bayesian update proposition, which uses the actual observed value.  We then show that the posterior probability (or probability density) of a theory is equal to the prior probability (or probability density) multiplied by the ratio of the $p$-value for the data obtained, given that theory, to the mean $p$-value --- averaged over all theories weighted by their prior probabilities. Thus, we provide frequentist answers to Bayesian questions. Our result is generic --- it does not rely on restrictive assumptions about the situation under consideration or specific properties of the likelihoods or the priors. 

\end{abstract}

\maketitle

\newpage 

\tableofcontents

\section{Introduction}
The two major statistical schools of thought, namely the frequentist and the Bayesian methods, have been broadly used in many scientific situations for probabilistic inference. The advantages and disadvantages of each competing approach have been debated for a long time. On the other hand, the connection between the two approaches, if any, is also of great interest. Despite the efforts put in this regard, such connections have been studied, to our knowledge, only in some limited situations.  We briefly mention a few efforts in this direction but do not necessarily try to be comprehensive.

J. O. Berger and Sellke~\cite{Berger1987} studied the point null hypothesis\footnote{A point null hypothesis is a hypothesis $H_0$ with a sharp prediction about a continuous parameter, i.e. $H_0: \mu=\mu_*$. } test. Considering a large class of prior probability densities, they showed that the $p$-value and the posterior probability for the null hypothesis can be very different. Thus, they concluded that the two methods cannot be reconciled.  See also an earlier paper by Dickey \cite{Dickey} with a similar conclusion.

Inspired by (but in contrast with) Ref.~\cite{Berger1987}, Casella and R. L. Berger~\cite{Casella1987} studied the connection between the $p$-value and the posterior probabilities for the case that we call threshold testing.\footnote{We use the phrase ``threshold testing" to describe a test of a hypothesis of the form $H_0: \mu<\mu_*$ for some particular value $\mu_*$. Note that in Ref.~\cite{Casella1987} this test is called ``one-sided", while we use the phrase one-sided in a different way in our Sec.~\ref{sec:one-sided}.} Since the posterior probability depends on the choice of a prior probability while the $p$-value does not, Ref.~\cite{Casella1987} compares the $p$-value with the upper or lower bound of the posterior probabilities for several classes of prior probability densities.  Restricting the analysis to symmetric, location likelihood densities\footnote{ A likelihood density $p(\ms|\mo)$ --- the probability density for obtaining the measured value $\ms$, given a theory described by the parameter $\theta$ --- is called a location likelihood density if it has the form $p(\ms|\mo)= f(|\mo-\ms|)$.} which have a monotone likelihood ratio,\footnote{A probability density $p(\ms|\mo)$ has a monotone likelihood ratio if the ratio $p(\ms|\mo_1)/p(\ms|\mo_2)$ is non-decreasing in $\ms$ whenever $\mo_1>\mo_2$ \cite{Karlin1956}.} Ref.~\cite{Casella1987} shows that the $p$-value is equal to the infimum of the posterior probability for specific classes of prior probability densities and greater than or equal to the infimum of the posterior probability for other specific classes of prior probability densities.  While this result does not show equivalence between the two methods, it nonetheless establishes a relationship between the two methods in certain situations. 

Micheas and Dey~\cite{Micheas} did a similar analysis to Ref.~\cite{Casella1987}, but studied one-sided scale parameter testing.\footnote{A scale parameter test is one for which the likelihood density belongs to a scale parameter family, i.e., it satisfies $p(\ms|\mo)=\frac{1}{\mo}f(\ms/\mo)$.} The Bayesian variables that they considered are the prior and the posterior predictive $p$-values.\footnote{The prior predictive $p$-value is defined by $\int P(\mu \geq \ms|\mo) p(\mo)d\mo$, where $p(\mo)$ is the prior probability density and $\ms$ is the measured value. The posterior predictive $p$-value is defined analogously with the prior in the integrand replaced with the posterior probability density.} For several classes of prior densities, they proved that the infimum of the prior and the posterior predictive $p$-values are both equal to the $p$-value. 

In this paper, we formulate a Bayesian question for which the answer may be given in terms of frequentist quantities.  In contrast to the papers cited above, we consider a more general class of questions that a Bayesian analyst could ask. A key difference between the frequentist approach and the standard Bayesian approach is that the frequentist's $p$-value is the probability, according to some theory, that a typical measurement would give a result more ``extreme'' than the value that is actually measured. The standard Bayesian approach, however, depends only on the probability or the probability density, according to the theory, of obtaining the result that was actually obtained. To bridge this gap, we consider the possibility that a Bayesian analyst could choose to update his probabilities not by using directly the measured value, but instead by using the weaker statement that the measurement yielded a value at least as ``extreme'' as the measured value.  Using this freedom, we can construct an expression for posterior probabilities written in terms of $p$-values, that holds without any restrictive assumptions about the nature of the tests or the properties of prior probability densities or likelihood densities.  This expression gives the Bayesian analyst a way to understand the question to which the $p$-value is the answer.

\section{A general analysis}

In this section we review the Bayesian method, introduce some notation, and also obtain some general results that may be applied to specific examples that will be given in the following sections. The theories to be tested, which we denote by $\calT(\mo)$, are theories that attempt to predict the value of the measurable quantity $\mu$. More precisely, the theory $\calT(\mo)$ predicts a probability density function (PDF) $p(\mu|\mo)$ that $\mu$ is observed, given $\calT(\mo)$.  The parameter $\mo$ that labels the theories might be taken to be the mean, median, or mode of $p(\mu|\mo)$, but any parametrization can be used. The PDF $p(\mu|\mo)$ (which is also called the likelihood density) incorporates the possibly probabilistic nature of the prediction as well as any uncertainty due to the imprecision of the measurement.

A probability for a continuous variable is conventionally described by a PDF: $p(\tilde \mu) d \tilde{\mu}$ is the probability that the continuous variable $\mu$ lies between $\tilde \mu$ and $\tilde \mu+d\tilde \mu$. In discussing a hypothesis $H$ which has a non-zero probability, the probability is conventionally indicated by $P(H)$. In some cases a non-zero probability might be assigned to a particular value of a continuous parameter (e.g., we might assign a non-zero probability to the hypothesis that the magnitude of the cosmological dipole modulation is exactly zero \cite{Planck:2019evm,Akrami:2014eta}). In this case, the situation can be described by a non-zero probability $P(H)$, or we can use a probability density function $p(\mo)$ that contains a delta function such as $P_H \delta(\mo-\mu_H)$. In what follows we will discuss continuous variables in terms of probability density functions, assuming that any assignment of non-zero probabilities to particular values can be treated using delta functions.

We first obtain some general relations describing a situation in which a single real continuous parameter $\mu$ is measured. Suppose that a measurement is made yielding the data $D$, where we have in mind that $D$ represents a proposition about the outcome with a non-zero probability. For example, $D$ could be the statement that the measured variable $\mu$ lies in a specified bin, or that $\mu$ is greater than some particular value.  The Bayes formula for the posterior probability density $p(\mo|D)$, after a measurement yielding the data $D$, is given by:
\ba
\label{eq:Bayes}
p(\mo|D)=\dfrac{P(D|\mo) p(\mo)}{\langle P(D)\rangle_{\text{pr}}},
\ea 
where $p(\mo)$ is the previously assumed probability density for $\mo$ (i.e., the so-called prior probability density), $P(D|\mo)$ is the probability of $D$ given the theory $\calT(\mo)$, and
\ba 
\langle p(D)\rangle_{\text{pr}}\equiv \int P(D|\mo) p(\mo) d\mo.
\ea 
Note that the denominator of Eq.~\eqref{eq:Bayes} is fixed by the requirement that $p(\mo|D)$ is normalized.

In the frequentist approach, it is the so-called $p$-value that is usually computed. The $p$-value is the probability, under the null hypothesis, that a measurement will yield a value that is at least as ``extreme" as the actual measurement. The meaning of extremity depends on the situation under consideration, as we shall discuss in the subsequent sections. For now, we obtain general relations that can be applied to any definition of extremity, allowing for the possibility that the definition of extremity might depend on the theory $\calT(\mo)$.  Denoting the measured value by $\ms$, we denote the set of extreme values of $\ms$ as follows:
\ba 
\label{eq:Uextreme}
U_\mo(\ms)&\equiv & \{ \mt\, |\, \text{$\mt$ is at least as extreme as $\ms$, under $\calT(\mo)$} \}.
\ea 
Then, the $p$-value for the measurement $\ms$ under the theory $\calT(\mo)$ can be defined by
\ba 
\label{eq:pval}
p_{\text{val},\, \mo}(\ms) \equiv \int_{U_\mo (\ms)} p(\mt|\mo) d\mt.
\ea 
Note that, as mentioned, the conventional practice for a frequentist is to compute the $p$-value for the null hypothesis. That is, denoting the null hypothesis by $\calT(\theta_s)$ --- where $s$ stands for {\it{standard}} --- the standard $p$-value is given by
\ba 
\label{eq:pval0}
p_\text{val}(\ms) \equiv p_{\text{val}, \, \theta_s}(\ms).
\ea 
Thus, Eq.~\eqref{eq:pval} is a generalization of the standard $p$-value (and we distinguish the generalized $p$-value, Eq.~\eqref{eq:pval}, from the traditional one, Eq.~\eqref{eq:pval0}, by the subscript $\mo$ on the former).  It is, however, a minor generalization, as a generalized $p$-value for one fequentist might be the standard $p$-value for another frequentist who is considering a different null hypothesis.  This generalization will allow us to relate the Bayesian to the frequentist. 

Now we consider what question a Bayesian might ask, if he is trying to understand the language of the frequentists. In a standard Bayesian treatment, $p(\mo)$ would be updated using the data that $\mu$ has been measured with the result $\ms$. However, to make contact with the frequentist approach, a Bayesian can instead use only the weaker statement that the measured value is at least as extreme as $\ms$. That is, the Bayesian can update the probabilities assigned to theories using the proposition $D'_\mo(\ms)$,
\ba 
\label{eq:D'_def}
D'_\mo(\ms) \equiv \text{The hypothesis that the measured value $\mu$ is in the set} \, U_\mo (\ms),
\ea 
which we refer to as the {\it extremity proposition.} This immediately implies that
\ba
\label{eq:pval_likelihood}
p_{\text{val},\, \mo}(\ms) = P(D'_\mo(\ms) |\mo).
\ea
As a consequence, Bayes' theorem (Eq.~\eqref{eq:Bayes}) becomes
\ba 
\label{eq:main}
p\big(\mo | D'_\mo(\ms) \big) = \dfrac{ p_{\text{val},\, \mo}(\ms) p(\mo)}{\langle p_{\text{val}}(\ms) \rangle_{\text{pr}}} \, ,
\ea 
where
\ba 
\langle p_{\text{val}}(\ms) \rangle_{\text{pr}}\equiv \int p_{\text{val},\, \mo}(\ms) p(\mo) d\mo.
\ea 
Eq.~\eqref{eq:main} is our key relation: as a result of the specific question that was asked, the posterior probability density can be written in terms of the prior probability density and the generalized $p$-value. 

In the following sections we will examine how Eq.~\eqref{eq:main} works in specific situations. Different situations lead to different natural definitions of extremity, which in turn lead to more explicit expressions for $p\big(\mo | D'_\mo(\ms) \big)$. 

\section{One-sided hypothesis testing} 
\label{sec:one-sided}
\begin{figure}
	\includegraphics[scale=.6]{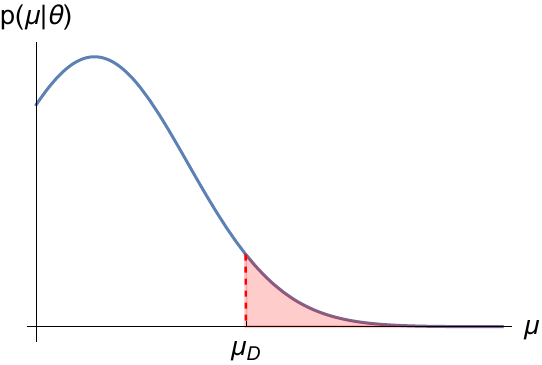}
	\hspace{1.2cm}
	\includegraphics[scale=.6]{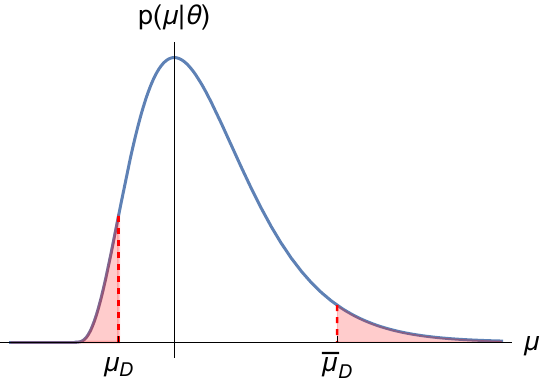}
	\caption{Illustration of the generalized $p$-value, for a given theory $\calT(\mo)$,  in one-sided (left) and two-sided (right) testing.  $p(\mu|\mo)$ is the  probability distribution for measuring  $\mu$  according to the theory $\calT(\mo)$. The generalized $p$-values are given by the areas of the shaded regions. The right graph is drawn under the assumption that $P(\mt \leq  \ms|\mo)<1/2$. $\bar \mu_D$ is defined via Eq.~\eqref{eq:musprime_def}, so that the area of the two shaded regions are equal and the entire shaded region depicts the generalized $p$-value.}
	\label{fig:pvalue_1}
\end{figure}

We first consider one-sided hypothesis testing.  An example could be the measurement of a chemical concentration, such as the concentration of lead in drinking water. We denote the concentration of lead by a continuous variable $\mu$. We again consider a one parameter family of theories $\calT(\mo)$ with $\calT(\theta_s)$ taken as the null hypothesis. In this example, it is natural to take $\mo$ to be the actual concentration of lead in the water, but in general we view $\mo$ as an arbitrary parametrization. In this example, the spread of $p(\mu|\mo)$ would be determined entirely by experimental uncertainties.  For the case of one-sided hypothesis testing, extremity is naturally defined by
\ba
\label{eq:U_onesided}
U_{\mo}(\ms) \equiv \left\{\mt \, |\, \mt \geq \ms \right\} .
\ea
According to Eq.~\eqref{eq:U_onesided}, being more extreme means being farther out into the tail.  In this case the definition of extremity does not depend on $\mo$, but we still use the notation $U_{\mo}(\ms)$ for consistency with our general formula. Using this definition, according to Eq.~\eqref{eq:pval_likelihood}, the generalized $p$-value is given by
\ba 
\label{eq:pval_onesided}
p_{\text{val},\, \mo}(\ms)= P(\mt \geq \ms|\mo) =\int_{\ms}^\infty p(\mt|\mo) d\mt .
\ea 
See Fig.~\ref{fig:pvalue_1} (left) for a visualization of the generalized $p$-value. Eq.~\eqref{eq:pval_onesided} can be used with Eq.~\eqref{eq:main} to obtain the posterior probability density for $\mo$, giving
\ba 
\label{eq:post_onesided}
p\big(\mo | D'_\mo(\ms) \big) =\dfrac{p(\mo) \int_{\ms}^\infty p(\mt|\mo) d\mt}{ \int_0^{\infty} \left[ p(\moh) \int_{\ms}^\infty p(\mt|\moh) d\mt\right] d\moh}.
\ea 
In the above equation we assumed that $\theta=0$ is the lower bound, as would be the case if $\theta=0$ corresponds to zero concentration. Alternatively, $\theta$ could for example be identified with log(concentration), in which case the lower limit of the integral over $\theta'$ in Eq.~\eqref{eq:post_onesided} would be changed to $-\infty$. In this case, it would still be natural to take Eq.~\eqref{eq:U_onesided} as the definition of extremity, since small values of the concentration (corresponding to large negative values of log(concentration)) are not concerning.

\section{Two-sided hypothesis testing}
We next consider two-sided, not necessarily symmetric, hypothesis testing.  To obtain some intuition about what extremity might mean in this situation, let us consider the case where $P(\mt \leq \ms|\mo)<1/2$, i.e., the measured value is located on the left half of the distribution. (The other possibility can be studied analogously.) Clearly, any value of $\mu$ less than or equal to $\ms$ can be considered at least as extreme as $\ms$ (just like the one-sided case). However, for a two-sided distribution, there are values of $\mu$ on the other side which can be viewed as equally extreme. A natural choice is to define an extreme region on the right with equal probability to the extreme region on the left. That is, define ${\bar \mu}_D$ such that
\ba 
\label{eq:musprime_def}
P(\mt \geq {\bar \mu}_D|\mo) = P(\mt \leq \ms | \mo )\, ,
\ea 
and then define the set of extreme values as $U_\mo(\ms)=\left \{\mt\, |\, \mt \geq {\bar \mu}_D \, \, \text{or} \, \, \mt \leq \ms \right \}$. Consequently, from Eq.~\eqref{eq:pval}, the generalized $p$-value is $p_{\text{val},\, \mo}(\ms)=P(\mt \leq \ms|\mo)+P(\mt \geq {\bar \mu}_D| \mo )=2P(\mt \leq \ms|\mo)$.  Dropping the simplifying assumption that $P(\mt \leq \ms|\mo)<1/2$, these results generalize to
\ba 
\label{eq:U_two_sided}
U_\mo(\ms)&=&\left \{\mt\, |\, \mt \geq \max \{ \ms, {\bar \mu}_D \}\, \, \text{or} \, \, \mt \leq \min \{ \ms, {\bar \mu}_D\} \right \}\, ,
\\
\label{eq:pval_twosided}
p_{\text{val},\, \mo}(\ms)&=& 2 \min \{P(\mt \geq \ms|\mo) , P(\mt \leq \ms|\mo) \}\, ,
\ea 
where ${\bar \mu}_D$ is still defined by Eq.~\eqref{eq:musprime_def}. See Fig.~\ref{fig:pvalue_1} (right) for a visualization of the generalized $p$-value for two-sided testing. Eqs.~\eqref{eq:U_two_sided} and \eqref{eq:pval_twosided} can then be used to obtain the posterior probability density when $D'_\mo(\ms)$ is used in Bayes' theorem, according to Eq.~\eqref{eq:main}. As a final remark, note that from Eq.~\eqref{eq:musprime_def} it follows that $ p_{\text{val},\,\mo}(\ms)= p_{\text{val},\, \mo}({\bar \mu}_D)$, $U_\mo(\ms)=U_\mo({\bar \mu}_D)$ and $D'_\mo(\ms)=D'_\mo({\bar \mu}_D)$. 

\section{Threshold testing}
Finally, we consider the situation in which one is not concerned about the actual value of a continuous parameter but instead asks whether the parameter exceeds a particular threshold. Following the example of one-sided testing discussed in Sec.~\ref{sec:one-sided}, an example of threshold testing is to determine whether the concentration of lead in a drinking water supply exceeds a specific safety guideline.  The null hypothesis $H_0$ can be described by the inequality $\mo <\mth$, where $\mo$ denotes the actual concentration of lead in drinking water while $\mth$ denotes the safety guideline threshold. Our goal would be to obtain the posterior probability for $H_0$ following a measurement of $\mu$ yielding the value $\ms$, using only the proposition that the measured value is more ``extreme'' than $\ms$. Note that even though we are only interested in the posterior probability for $H_0$, we still assume that the prior PDF for $\mo$ has been specified. In this situation, extremity is naturally defined by
\ba 
\label{eq:U_threshold}
U_\mo(\ms) \equiv \{\mt \, |\, \mt \geq \ms \}\, ,
\ea 
which is identical to Eq.~\eqref{eq:U_onesided} for one-sided testing.  Although $\mu\geq 0$ in our example, we might alternatively --- as discussed in Sec.~\ref{sec:one-sided} --- use $\mu$ to denote, e.g., the logarithm of the concentration, which would then have no lower bound.  Eq.~\eqref{eq:U_threshold} would still be the natural definition of extremity, because the other end of the range, $\log\text(\rm concentration)$ large and negative, is consistent with the null hypothesis.  As in Sec.~\ref{sec:one-sided}, we will write the equations for this case under the assumption that $\mo=0$ is the lower bound. If $\theta$ is unbounded, all the integrals over $\mo$ in the equations that follow should be modified by changing the lower limits from $0$ to $-\infty$.

Using Eq.~\eqref{eq:main}, we again obtain Eq.~\eqref{eq:post_onesided}, the same as for one-sided testing. However, here we need to take one further step to obtain the posterior probability for $H_0$:
\ba 
\label{eq:post_threshold_int}
P \big(H_0|D'_\mo(\ms)\big)=\int_{0}^{\mth} p \big(\mo|D'_\mo(\ms)\big) d\mo =\dfrac{\int_{0}^{\mth}\left[ p(\moh) \int_{\ms}^\infty p(\mt|\mo) d\mt\right] d\moh}{\int_{0}^{\infty} \left[ p(\moh) \int_{\ms}^\infty p(\mt|\mo) d\mt \right] d\moh}.
\ea 
This result may be expressed in a more intuitive way by writing the prior probability of $H_0$ as $P(H_0)= \int_{0}^{\mth} p(\mo)d\mo$, and denoting the expected generalized $p$-value according to the null hypothesis by
	\ba 
	\langle p_\text{val}(\ms) \rangle_{\text{pr},{H_0}} \equiv  \dfrac{1}{P(H_0)} \int_{0}^{\mth} p_{\text{val},\, \mo}(\ms) p(\mo) d\mo.
	\ea 
     Then, Eq.~\eqref{eq:post_threshold_int} can be rewritten as
\ba 
P \big(H_0|D'_\mo(\ms)\big) = \dfrac{ \langle p_\text{val}(\ms) \rangle_{\text{pr}, {H_0}}}{\langle p_\text{val}(\ms)\rangle_{\text{pr}}} P(H_0).
\ea 
Thus, given the data $D'_\mo(\ms)$, the posterior probability for the null hypothesis is proportional to the expected generalized $p$-value according to the null hypothesis.

\section{Conclusion}

Both the Bayesian and frequentist approaches to assessing the implications of measurements remain in common use, so it is certainly worthwhile to explore the relation between them.  They differ somewhat in their precise goals.  The Bayesian approach is a direct attempt to answer the fundamental question that a scientist faces: based on some set of experiments, how should one assess the probability that various alternative theories or hypotheses might be true? The frequentist approach, on the other hand, addresses directly only the probability that the observed data could arise, given a particular theory, but makes no attempt to assign a probability to the theory itself.

Advocates of the frequentist approach sometimes fault the Bayesian approach for its reliance on assumed prior probabilities, which are ultimately subjective.  The Bayesian might reply that the need for priors cannot be rationally avoided.  For example, if I look out my window and discover that the ground and trees are wet, I might consider two theories to explain this: (a) aliens from Alpha-Centauri landed and sprayed water over everything, or (b) it rained.  The two theories would account for the observations about equally well, but undoubtedly most of us would give much higher odds to theory (b).  In Bayesian language, theory (b) would be assiged a much higher prior.  Priors can play a major role in our reasoning, whether we are formally using Bayes' theorem, or just relying on common sense.

Given that the Bayesian insists on trying to assign actual probabilities to theories, she might well ask whether $p$-values can be of any use in this regard.  A key obstacle to drawing connections between the two approaches is that the standard Bayesian and frequentist approaches base their conclusions on different quantities.  In the standard Bayesian approach, the probabilities assigned to different theories are updated on the basis of a new measurement, making use of the ``likelihood,'' which is the probability or probability density of obtaining the result that was actually obtained, according to each theory.  The standard frequentist approach, however, is to begin by defining some notion of what it means for a result to be ``extreme''.  The plausibility of any theory is then assessed by the ``$p$-value,'' which is the probability, according to the theory, that a measurement of the type performed would give a result {\em at least as extreme} as the result that was actually obtained. 

To bridge this dichotomy in a way that does not depend on finding special cases where the likelihoods and $p$-values are related, here we consider the possibility that the Bayesian can choose to update her probabilities on the basis of a different input.  Instead of using the actual data from an experiment, the Bayesian could update her probabilities using only the weaker {\it extremity proposition}, the statement that the results of the experiment were at least as extreme as the data that was obtained.  The approach is still fully Bayesian, using new information to update probabilities assigned to theories.  With this modification of the approach, we have found that the Bayesian can update probabilities by directly using $p$-values, or at least ``generalized'' $p$-values.  By generalized $p$-values, we mean $p$-values calculated for arbitrary theories, rather than just the null hypothesis.

With this modification of the update proposition, we have found that the posterior probability assigned to theory $\calT(\theta)$ is expressed by Eq.~\eqref{eq:main}, which is our main and generic result.  In words, the posterior probability for $\calT(\theta)$ is equal to the prior probability, multiplied by the ratio of the $p$-value for the data obtained, given the theory $\calT(\theta)$, to the mean $p$-value for the data obtained, averaged over all theories weighted by their prior probabilities.

This relationship becomes particularly simple if one considers the Bayes factor for two theories, $\calT(\theta_i)$ and $\calT(\theta_j)$, which is generally defined by
\ba
B_{\theta_i \theta_j}(D) \equiv \dfrac{P(D|\theta_i)}{P(D|\theta_j)} =
\frac{p(\theta_i|D) / p(\theta_i)}{p(\theta_j|D) /
     p(\theta_j)} \ .
\ea
If instead of updating with the proposition $D$, the full data, the Bayesian analyst uses the weaker {\it extremity proposition} $D'_{\theta}(\ms)$ --- the proposition that the observed data was at least as extreme as the data actually observed --- then there is a simple relationship with $p$-values:
\ba
B_{\theta_i \theta_j}(\ms) \equiv \dfrac{P\big(D'_{\theta_i}(\ms)|\theta_i\big)}{P\big(D'_{\theta_j}(\ms)|\theta_j\big)} = \dfrac{ p_{\text{val},\, \theta_i}(\ms)}{ p_{\text{val},\, \theta_j}(\ms)}\, ,
\ea 
where the final equality follows from Eq.~\eqref{eq:pval_likelihood}. Thus, by using the modified update proposition $D'_{\theta}(\ms)$, the Bayes factor is precisely equal to the ratio of $p$-values.  So, in this context, the Bayesian can clearly state the question to which the $p$-value (or at least the ratio of $p$-values) is the answer.

\end{document}